\documentclass{PoS}

\title{Strangeness and glue in the nucleon from lattice QCD }

\ShortTitle{Strangeness and glue in the nucleon from lattice QCD }

\author{\speaker{Takumi Doi},
{Mridupawan Deka},
{Shao-Jing Dong},
{Terrence Draper},
{Keh-Fei Liu}
and
{Devdatta Mankame} \\
Department of Physics and Astronomy,
University of Kentucky, Lexington KY 40506, USA \\
E-mail: 
\email{doi@pa.uky.edu},
\email{mpdeka@pa.uky.edu},
\email{s.j.dong@uky.edu},
\email{draper@pa.uky.edu},
\email{liu@pa.uky.edu},
\email{dmmank2@uky.edu}
}

\author{Nilmani Mathur\\
Department of Theoretical Physics, 
Tata Institute of Fundamental Research,
Mumbai 40005, India \\
E-mail: \email{nilmani@theory.tifr.res.in}}

\author{Thomas Streuer\\
Institute for Theoretical Physics, 
University of Regensburg, 93040 Regensburg, Germany \\
E-mail: \email{thomas.streuer@physik.uni-regensburg.de}}

\author{${\rm \chi QCD}$ collaboration}

\abstract{

We study the strangeness contribution to nucleon matrix elements
using $N_f=2+1$ dynamical clover fermion configurations generated
by the CP-PACS/JLQCD collaboration.
In order to evaluate the disconnected insertion (DI),
we use the Z(4) stochastic method, along with 
unbiased subtraction from the hopping parameter expansion which reduces
the off-diagonal noises in the stochastic method.
Furthermore, we find that using many nucleon sources
for each configuration is effective in improving the 
signal. Our results for the quark contribution to the first moment $\xnuc_q$
in the DI,
and the strangeness magnetic moment show that the statistical errors are
under control with these techniques.
We also study the gluonic contribution to the nucleon
using the overlap operator
to construct the gauge field tensor, $F_{\mu\nu}$.
The application to the calculation of first moment, $\xnuc_G$,
gives a good signal in quenched lattice QCD.

}

\FullConference{The XXVI International Symposium on Lattice Field Theory\\
		 July 14-19 2008\\
		 Williamsburg, Virginia, USA}

\newcommand{\bra}{\langle}
\newcommand{\ket}{\rangle}
\newcommand{\braket}[1]{\bra #1 \ket}

\newcommand{\xnuc}{\langle x \rangle}
\newcommand{\xxnuc}{\langle x^2 \rangle}

\begin{document}

\section{Introduction}

The understanding of the structure of nucleon 
is one of the central issues in hadron physics.
For instance, the parton distribution functions (PDFs)
have been studied extensively, and the observation
of scaling violation in PDFs 
provides the cradle for the 
establishment of the fundamental theory, QCD.
Yet, there exist many unresolved questions for the
structure of the nucleon. The EMC experiments~\cite{emc} and subsequent
experiments show that quark spin carries only $\sim$ 30\% of
the total spin of the nucleon. Consequently, one concludes 
that the remaining $\sim$ 70\%
should be carried by the quark orbital contribution and
the glue.
Direct quantitative
identification of each contribution has been undertaken by experiments.
The strangeness contribution to the nucleon structure is
also under intensive study experimentally.
%For instance, 
%with regard to the strangeness magnetic moment
%of nucleon, even the overall sign is still uncertain.
%the assymmetry in strange and anti-strange quark distribution

Under these circumstances, it is desirable 
to provide definitive quantitative results
using the lattice QCD method.
%directly from QCD, 
%and the lattice QCD, the 
%first principle calculation of QCD, is a 
%suitable method for that.
%
In fact, there are many lattice QCD studies
of the nucleon structure~\cite{lat08:zanotti}. 
However, most of 
these calculations are limited to the so-called
``connected insertion (CI)'', and there have been few calculations
considering ``disconnected insertion (DI)''%
~\cite{smm:ky_quenched,angular:ky_quenched,smm:randy}.
Although the calculation of DI is known to be a very
difficult problem, we emphasize that DI are
related to rich physics, e.g., only DI
consists of the the strangeness contribution to the nucleon.
% because there is no valence strange quark
%in the nucleon.
We shall describe our methodology to obtain the
signal effectively in DI calculation.
We also note that using $N_f=2+1$ dynamical configurations 
could be essential
for small quark masses.
%because strangeness in nucleon can emerge
%only through the dynamical strangeness quark loop, in principle.
In this proceeding, we present the study of the DI part for the 
quark contribution to 
the first moment of the nucleon, $\braket{x}_q$,
whereas the study of the CI part is presented elsewhere~\cite{lat08:dev}.

Another very important, but often not calculated component,
is the glue contribution to the nucleon structure.
This is because the straight-forward calculation
using the standard link variables is known to 
%provide
yield
%just 
very noisy signal~\cite{glue:gockeler}.
We have proposed~\cite{gtensor} to use the overlap operator ($D_{ov}$) to 
overcome this problem,
and we show
%in order to suppress the high-frequency modes in the
%gluon operator.
%We will show our 
the first application of ${\rm Tr^s}[\sigma_{\mu\nu}D_{ov}]$ to the gluonic first 
moment, $\xnuc_G$.
% is given in this proceeding.

\section{Formalism and simulation parameters}

We use the $N_f=2+1$ dynamical clover fermion 
with renormalization group improved gauge configurations
generated by CP-PACS/JLQCD collaboration~\cite{conf:tsukuba2+1}.
We use the $\beta=1.83$ configurations 
with the lattice size of $V = L^3 \times T = 16^3\times 32$,
for which the lattice unit is $a^{-1}=1.62 {\rm GeV}$
and physical spatial size is $(2{\rm fm})^3$.
The hopping parameters for light (u,d) quarks are 
$\kappa =$ $0.13825$, $0.13800$ and $0.13760$,
which correspond to $m_\pi =$ $610$, $700$ and $840$ ${\rm MeV}$,
respectively, and the hopping parameter for strange quark 
is fixed to be $\kappa_s = 0.13760$. We perform the calculation
only for the dynamical quark mass points.
For each quark masses, about 800 configurations are used.

We also perform the complementary study
using the Wilson fermion with Wilson gauge action
in the quenched approximation.
We generate 500 configurations of $16^3 \times 24$ lattice
at $\beta=6.0$,
where  the lattice spacing is $a^{-1} = 1.74 {\rm GeV}$
and the physical spatial size is $(1.8{\rm fm})^3$~\cite{angular:ky_quenched}.
The calculation is performed with three light quark hopping parameters 
%We use three light quark hopping parameters 
of $\kappa = 0.1540, 0.1550, 0.1555$,
which correspond to $m_\pi = 480-650 {\rm MeV}$,
with strange quark hopping parameter fixed to $\kappa_s = 0.1540$.

The nucleon matrix elements can be calculated 
by taking the ratio of three point function $\Pi^{\rm 3pt}_{\cal O}$
to two point function $\Pi^{\rm 2pt}$,
\begin{eqnarray}
\Pi^{\rm 3pt}_{\cal O}(\vec{p},t_2;\ \vec{q},t_1;\ \vec{p'},\vec{x}_0,t_0)
&=& \sum_{\vec{x_2},\vec{x_1}}
e^{-i\vec{p}\cdot\vec{x}_2}
e^{+i\vec{q}\cdot\vec{x}_1}
\braket{0|{\rm T}\left[
J_N(\vec{x}_2,t_2) {\cal O}(\vec{x}_1,t_1) \bar{J}_N(\vec{x}_0,t_0)
\right] |0} ,\\
\Pi^{\rm 2pt}(\vec{p},t;\ \vec{x}_0,t_0) 
&=& \sum_{\vec{x}} e^{-i\vec{p}\cdot\vec{x}}
\braket{0|{\rm T}\left[
J_N(\vec{x},t) \bar{J}_N(\vec{x}_0,t_0)
\right] |0} ,
\end{eqnarray}
where ${\cal O}$ is an appropriate operator 
for the matrix element of concern
and
$J_N$ is the nucleon interpolating field.

The calculations of three point functions of DI
involve evaluation of both of the two point function part 
and the quark loop part.
Because the latter requires all-to-all propagators, for which 
the straightforward calculations are practically impossible, 
we use the stochastic method~\cite{DI:noise} as follows
\begin{eqnarray}
{\rm Tr}[\Gamma D^{-1}] = 
\lim_{L\rightarrow\infty} \frac{1}{L} 
\sum_{l=1}^L \eta^{l\,\dag} \Gamma D^{-1} \eta^l,
\qquad
\lim_{L\rightarrow\infty} \frac{1}{L} 
\sum_{l=1}^L \eta^{l\,\dag}_i \eta^l_j = \delta_{ij} , 
\end{eqnarray}
where $\Gamma$ is an arbitrary matrix and $\eta^l$ corresponds to the $l$-th noise.
In the practical calculation, this method introduces a variance,
because $L=N_{noise}$ is finite.
In order to reduce such off-diagonal error~\cite{DI:noise}, 
we use the unbiased subtraction
from the hopping parameter expansion (HPE)~\cite{DI:hpe}.
In our practical calculation, we adopt $Z(4)$ noises
in color, spin and space-time indices.
We take $N_{noise} = 300 (500)$ for each configuration
for full (quenched) QCD simulation, respectively, along with the use of HPE
up to the 4th order.

In the stochastic method, 
it is quite expensive
to achieve a good signal to noise ratio (S/N) by just increasing $N_{noise}$
because S/N improves with $\sqrt{N_{noise}}$.
In view of this, we use many nucleon sources $N_{src}$ 
in the evaluation of the two point function part for each configuration.
Because the calculations of quark loop and those of two point functions
are independent, this is expected to be an efficient way to increase statistics.
In fact, as we will show later, we observe that S/N improves almost ideally,
i.e., by a factor of $\sqrt{N_{src}}$.

For the study of glue contribution to the nucleon structure,
it is essential to find a suitable glue operator.
In fact, glue operators constructed from link variables
suffer from large fluctuations in high-frequency modes,
which causes %very 
poor S/N in the calculation~\cite{glue:gockeler}.
We propose~\cite{gtensor} to use the gauge field tensor
constructed from the overlap operator $D_{ov}$ as
\begin{eqnarray}
F_{\mu\nu}(x) 
= {\rm const.} \times {\rm Tr^s} [ \sigma_{\mu\nu} D_{ov}(x,x) ], 
\label{eq:F_ov}
\end{eqnarray}
where ${\rm Tr^s}$ corresponds to the trace in spinor space.
The advantage of this formulation is that 
the ultraviolet fluctuation is expected to be suppressed
due to the exponential local  nature of $D_{ov}$.
%so that the automatic smearing is achieved.
In order to estimate $D_{ov}(x,x)$, we again use
the stochastic method. In this case, we treat the color and spin 
indices exactly and space-time indices are diluted for two sites separation
on top of the even/odd dilution. Therefore, 
the minimal length to the next neighbor site amounts to
four hoppings away  (``taxi driver distance''=4). We use two Z(4) noises and take the average
between them
for each configuration.

\section{First moment of parton distribution}

Quark contribution to the first moment of 
the parton distribution in the nucleon, $\xnuc_q$,
can be obtained by using the following energy-momentum tensor operator~\cite{ji:spin97},
\begin{eqnarray}
T_{4i} &=&
(-1)*\frac{i}{4}
\left[
\bar{q} \gamma_4 \overrightarrow{D}_i q
+ \bar{q} \gamma_i \overrightarrow{D}_4 q
- \bar{q} \gamma_4 \overleftarrow{D}_i q
- \bar{q} \gamma_i \overleftarrow{D}_4 q
\right] ,
\end{eqnarray}
and by taking the following ratio of three point to two point function,
\begin{eqnarray}
\frac{
{\rm Tr}\left[
\Gamma_e\cdot
\Pi^{\rm 3pt}_{T_{4i}}(\vec{p},t_2;\ \vec{0},t_1;\ \vec{p},\vec{x}_0,t_0)
\right] \times e^{-i\vec{q}\cdot\vec{x}_0}
}
{
(2 p_i)\times
{\rm Tr} \left[
\Gamma_e\cdot \Pi^{\rm 2pt}_{}(\vec{p},t_2;\ \vec{x}_0,t_0)
\right] \times e^{+i\vec{q}\cdot\vec{x_0}}
}
=
\xnuc_q ,
\end{eqnarray}
where 
$
\Gamma_e \equiv 
\left(
\begin{array}{cc}
1 & 0 \\
0 & 0 \\
\end{array}
\right)
$
in spinor space (for the Dirac representation.)
In order to improve the S/N, we further take the summation
for the operator insertion time $t_1$ for the range 
$t_1 = [t_0+1, t_2-1]$,
where $t_0 (t_2)$ is the nucleon source (sink) time, respectively.

\begin{figure}[b]
\begin{center}
\begin{tabular}{cc}
\begin{minipage}{73mm}
\begin{center}
   \includegraphics*[width=0.6 \textwidth,angle=270]
   {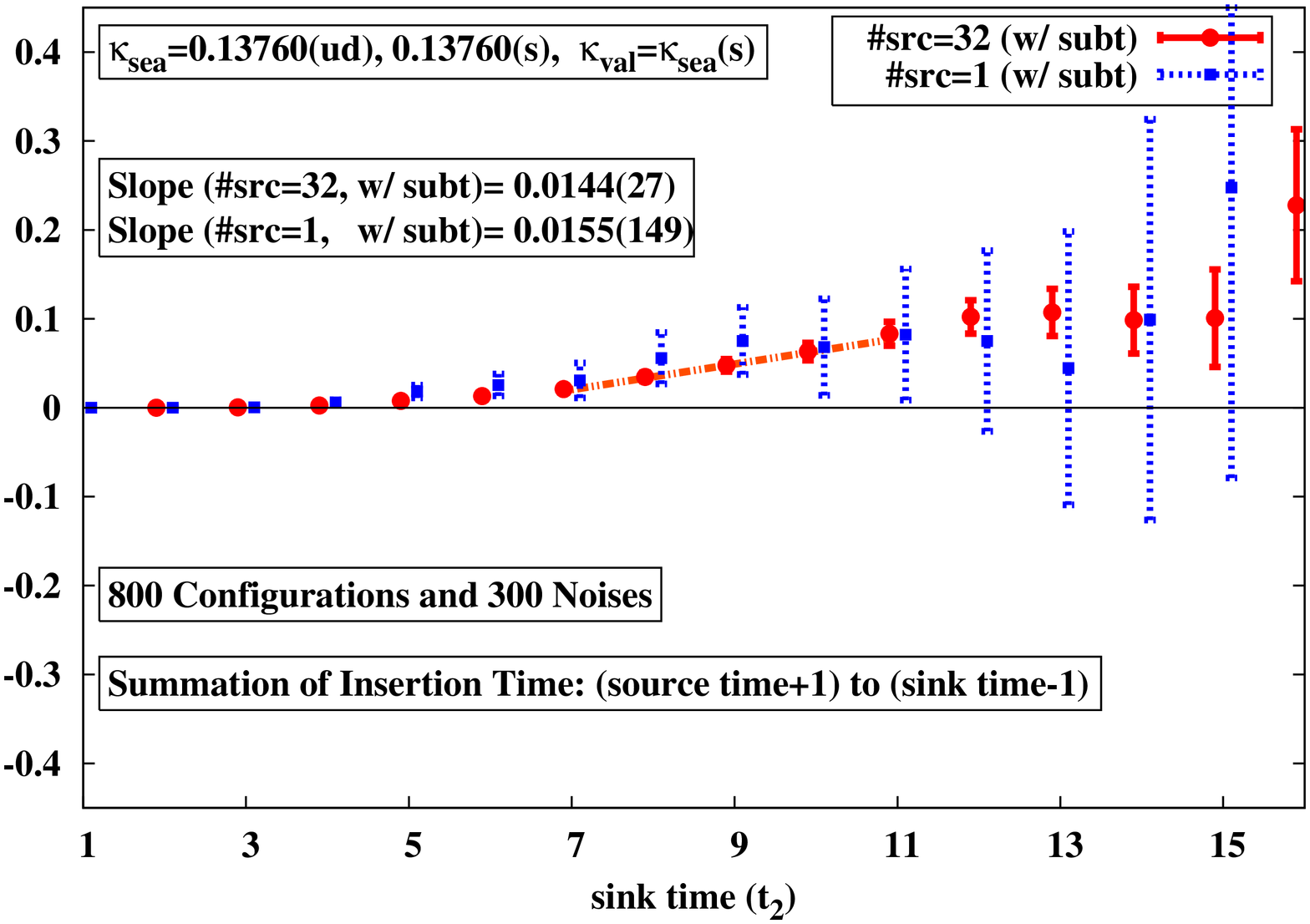}
 \caption{The ratio of three point to two point functions
plotted in terms of the nucleon sink time $t_2$
for dynamical configurations with $\kappa = 0.13760$.
The linear slope corresponds to $\xnuc_s$.}
 \label{fig:x_s}
\end{center}
\end{minipage}&
\begin{minipage}{73mm}
\begin{center}
\vspace*{-4mm}
   \includegraphics*[width=0.6 \textwidth,angle=270]
   {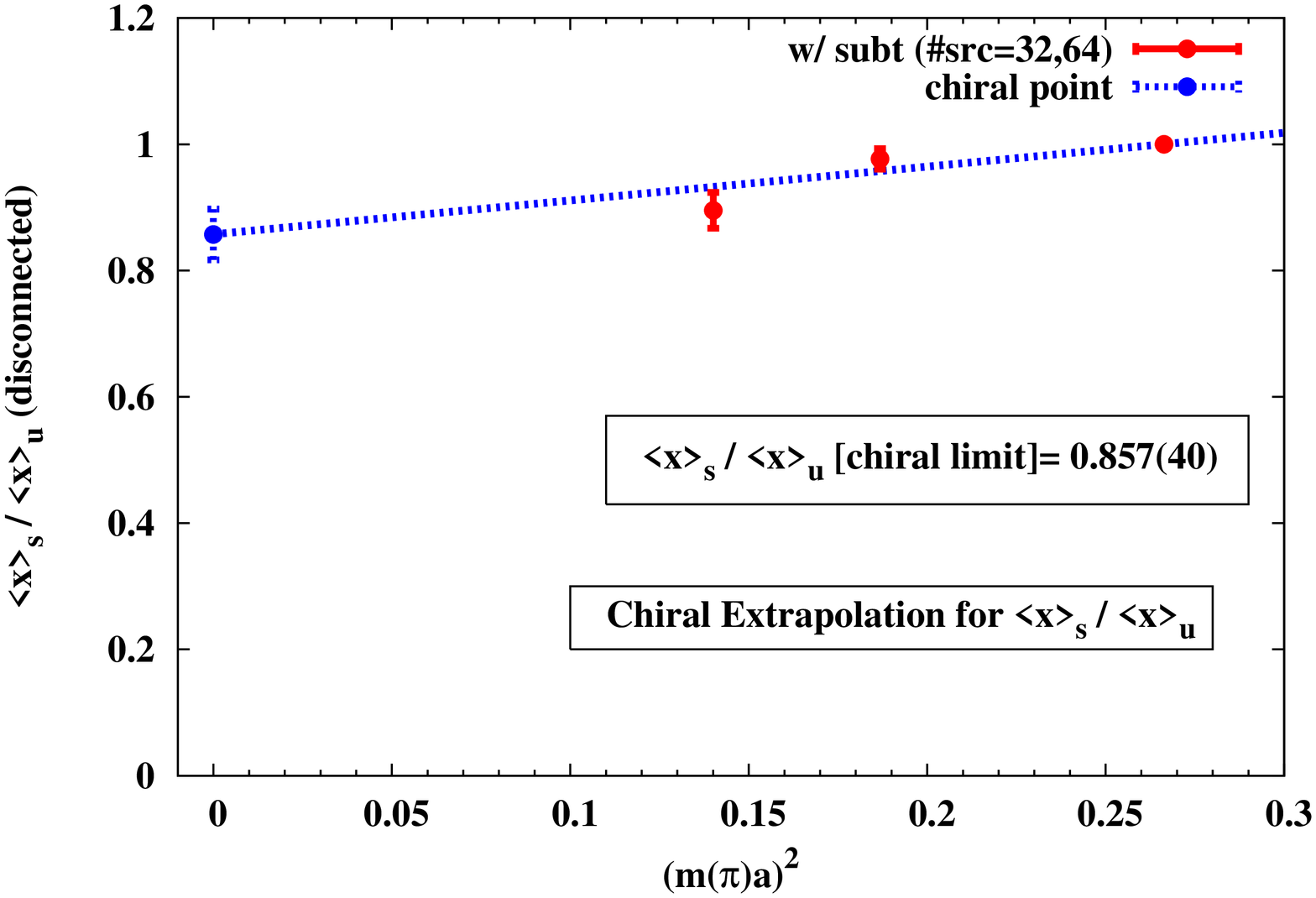}
 \caption{Ratio $\xnuc_s / \xnuc_{ud}(DI)$
plotted in terms of $m_\pi^2$ for dynamical configurations.
Blue line corresponds to the linear chiral extrapolation.
}
 \label{fig:x_ratio}
\end{center}
\end{minipage}
\end{tabular}
\end{center}
\end{figure}

In Fig. \ref{fig:x_s}, we plot the ratio of three point to two point
functions
for the strangeness, $\xnuc_s$, in terms of $t_2$.
Because of the summation of operator insertion time $t_1$, 
the linear slope corresponds to $\xnuc_s$.
Blue points denote the result for $N_{src}=1$ and red points
for $N_{src}=32$.
One can clearly see that increasing $N_{src}$ reduces the error
significantly (by about a factor of $\sqrt{N_{src}}$),
and a clear signal can be extracted from the $N_{src}=32$ data.

In Fig. \ref{fig:x_ratio}, we plot the ratio of
$\xnuc_s / \xnuc_{ud}(DI)$ for each quark mass.
By taking the linear chiral extrapolation
in terms of $m_\pi^2$, we obtain a preliminary result
\begin{eqnarray}
\xnuc_s / \xnuc_{u,d}(DI) = 0.857(40),
\end{eqnarray}
in the chiral limit.
Although it is necessary to consider the renormalization factor,
this result shows that the statistical error is well under control
and reliable lattice QCD calculation is possible even for the DI.

We perform the same calculation for the quenched case,
and the result using $N_{src}=16$  with perturbative renormalization 
is~\cite{x:deka}
\begin{eqnarray}
\xnuc_s / \xnuc_{u,d}(DI) = 0.88(7).
\end{eqnarray}

\begin{figure}[hbt]
 \centering
   \includegraphics*[width=0.3 \textwidth,angle=270]
   {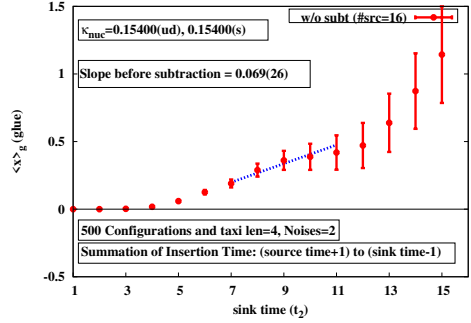}
 \caption{The ratio of three point to two point functions
plotted in terms of nucleon sink time $t_2$
for quenched configurations with $\kappa = 0.1540$.
The linear slope corresponds to $\xnuc_G$.}
 \label{fig:x_G}
\end{figure}

Gluonic contribution to the first moment, $\xnuc_G$, can be obtained
in the same way, by using the following energy-momentum tensor
\begin{eqnarray}
T_{4k} = - \sum_{i=1}^3 (+i) * ( F_{4i} F_{ki} ),
\end{eqnarray}
where $F_{\mu\nu}$ is constructed using the overlap operator 
as shown in Eq.~(\ref{eq:F_ov}).
In Fig. \ref{fig:x_G}, 
we plot the ratio of three point to two point functions in terms of the nucleon sink time $t_2$,
using 
the quenched QCD configurations.
As can be seen, we obtain a prominent linear signal for $\xnuc_G$.
It is quite encouraging that we obtain the signal with about three sigma accuracy.
With unbiased subtraction, this could be improved to larger than
four sigma accuracy.
We are also working on the deflation which could improve the S/N substantially.
Further investigation including the renormalization factor is in progress.

\section{Strangeness electric and magnetic form factors}

We use the point-split vector current as
\begin{eqnarray}
%
%J_\mu =
V_\mu =
\bar{q}(x) (1-\gamma_\mu) U_\mu(x) q(x+\mu)
- \bar{q}(x+\mu) (1+\gamma_\mu) U_\mu^\dag(x) q(x).
\end{eqnarray}
The advantage of the point-split operator is that
there is no additional renormalization factor 
to be considered,
because it is a conserved current.
Using this operator, the electric and magnetic form factors
can be obtained~\cite{smm:ky_quenched}
 by the following combination of three point and two point functions:
\begin{eqnarray}
\frac{
{\rm Tr}\left[
\Gamma_e\cdot
\Pi^{\rm 3pt}_{V_{\mu=4}}(\vec{0},t_2;\ \vec{q},t_1;\ -\vec{q},\vec{x}_0,t_0)
\right] \times e^{-i\vec{q}\cdot\vec{x}_0}
}
{
{\rm Tr} \left[
\Gamma_e\cdot \Pi^{\rm 2pt}_{}(\vec{q},t_1;\ \vec{x}_0,t_0)
\right] \times e^{+i\vec{q}\cdot\vec{x_0}}
}
\cdot
\frac{
{\rm Tr} \left[
\Gamma_e\cdot \Pi^{\rm 2pt}_{}(\vec{0},t_1;\ \vec{x}_0,t_0)
\right]
}
{
{\rm Tr} \left[
\Gamma_e\cdot \Pi^{\rm 2pt}_{}(\vec{0},t_2;\ \vec{x}_0,t_0)
\right]
} \nonumber \\
=
%\frac{1}{2\kappa}
\left[
F_1(-q_E^2) - \frac{E_q-m}{2m} F_2(-q_E^2)
\right]
\end{eqnarray}

\begin{eqnarray}
\frac{
{\rm Tr}\left[
\Gamma_k\cdot
\Pi^{\rm 3pt}_{V_{\mu=i}}(\vec{0},t_2;\ \vec{q},t_1;\ -\vec{q},\vec{x}_0,t_0)
\right] \times e^{-i\vec{q}\cdot\vec{x}_0}
}
{
{\rm Tr} \left[
\Gamma_e\cdot \Pi^{\rm 2pt}_{}(\vec{q},t_1;\ \vec{x}_0,t_0)
\right] \times e^{+i\vec{q}\cdot\vec{x_0}}
}
\cdot
\frac{
{\rm Tr} \left[
\Gamma_e\cdot \Pi^{\rm 2pt}_{}(\vec{0},t_1;\ \vec{x}_0,t_0)
\right]
}
{
{\rm Tr} \left[
\Gamma_e\cdot \Pi^{\rm 2pt}_{}(\vec{0},t_2;\ \vec{x}_0,t_0)
\right]
} \nonumber \\
=
%
%\frac{1}{2\kappa}
(-1)*\frac{1}{E_q+m}
\epsilon_{ijk} q_j^E \left( F_1(-q_E^2) + F_2(-q_E^2) \right) ,
\end{eqnarray}
where 
$
\Gamma_k \equiv 
\left(
\begin{array}{cc}
\sigma_k & 0 \\
0 & 0 \\
\end{array}
\right)
$,
$m$ denotes nucleon mass, and $E_q \equiv \sqrt{m^2 + \vec{q}^2}$.

\begin{figure}[hbt]
 \centering
   \includegraphics*[width=0.3 \textwidth,angle=270]
   {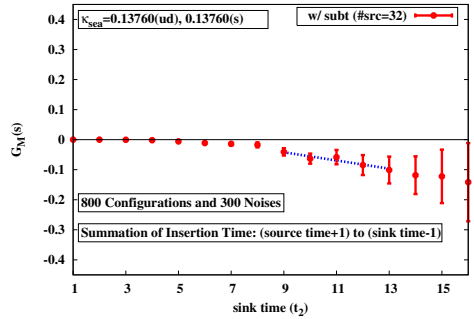}
 \caption{The ratio of three point to two point functions
plotted in terms of the nucleon sink time $t_2$
for dynamical configurations with $\kappa = 0.13760$.
The linear slope corresponds to the strangeness magnetic 
form factor at $\vec{q}^2 = (2\pi/L)^2$}.
 \label{fig:smm}
\end{figure}

In Fig. \ref{fig:smm}, 
we plot the ratio of three point to two point functions
in terms of the nucleon sink time $t_2$
where the linear slope corresponds to the signal
of strangeness magnetic form factor at $\vec{q}^2 = (2\pi/L)^2$.
%This result indicates that s.m.m. is small negative value,
%which is roughly consistent with our previous quenched study.
In order to obtain the final quantitative result for 
the strangeness magnetic moment,
detailed study for several $Q^2(-q^2)$ and chiral extrapolation
are needed and they are in progress.

\section{Summary}

We have studied the strangeness and gluonic contribution
to the nucleon matrix elements from lattice QCD.
The strangeness matrix elements have been studied
by the Z(4) stochastic method, 
with the unbiased subtraction from the hopping parameter expansion
in order to reduce the off-diagonal noises.
We have further taken many nucleon sources for each configuration,
and observed that this method is almost ideally effective to improve the
signal with modest cost.
Using $N_f=2+1$ dynamical clover fermion configurations,
we have analyzed the quark contribution to the first moment, $\xnuc_q$,
and the strangeness magnetic form factor,
and found that 
statistical errors are well under control
with these improvements.

The gluonic contribution for the first moment of nucleon, $\xnuc_G$,
has also been studied with the use of 
the overlap operator
to construct the gauge field tensor, $F_{\mu\nu}$,
and we have shown the effectiveness of this method 
by the explicit calculation at the quenched level.
%using the quenched lattice configurations.

Although there remain several sources of the systematic error,
such as the finite volume artifact,
discretization error, excited-states contamination and chiral extrapolation, 
we plan to investigate these issues
by using the configurations with larger volume and lighter 
quark masses with various lattice cut-offs.
The study for other matrix elements such as the angular momentum 
contribution to the nucleon spin is also in progress.

\acknowledgments

We thank the CP-PACS/JLQCD collaboration
for their configurations.
This work was supported  in part by 
U.S. DOE grant DE-FG05-84ER40154.
Research of N.M. is supported by Ramanujan Fellowship.
The calculation was performed on
supercomputers at Jefferson Laboratory
and the University of Kentucky.

\end{document}